\documentclass[apj]{emulateapj}

\usepackage{graphicx}
\usepackage{amsfonts}
\usepackage{amsmath}
\usepackage{amssymb}
\usepackage{amsthm}

\usepackage{textcomp}

\newcommand{\etal}{et~al.\ }

\newcommand{\pflux}{\hbox{phot~cm$^{-2}$~s$^{-1}$}}

\newcommand{\plumin}{\hbox{phot~s$^{-1}$}}

\newcommand{\msun}{\hbox{${M}_{\odot}$}}

\newcommand{\be}{\begin{equation}}
\newcommand{\ee}{\end{equation}}
\newcommand{\ba}{\begin{eqnarray}}
\newcommand{\ea}{\end{eqnarray}}

\newcommand{\simgt}{\lower 2pt \hbox{$\, \buildrel {\scriptstyle >}\over {\scriptstyle\sim}\,$}}
\newcommand{\simlt}{\lower 2pt \hbox{$\, \buildrel {\scriptstyle <}\over {\scriptstyle\sim}\,$}}
\newcommand{\ls}{\lower 2pt \hbox{$\;\scriptscriptstyle \buildrel<\over\sim\;$}}
\newcommand{\gs}{\lower 2pt \hbox{$\;\scriptscriptstyle \buildrel>\over\sim\;$}}

\newcommand{\fermi}{\emph{Fermi}}
\newcommand{\degree}{$^\circ$}

\newcommand{\jcap}{JCAP}

\slugcomment{Submitted to ApJ}
\shorttitle{Gamma-Ray Emission From Galaxy Clusters}
\shortauthors{Griffin, Dai, \& Kochanek}

\begin{document}

\def\arcsec{$^{\prime\prime}$}
\def\arcmin{$^{\prime}$}
\def\degr{$^{\circ}$}
\def\gam{$\gamma$}
\def\ddotu{$\ddot{\text{u}}$}
\def\deg{$^{\circ}\!\!.$}

\title{New Limits On Gamma-Ray Emission From Galaxy Clusters}

\author{Rhiannon D. Griffin\altaffilmark{1}, Xinyu Dai\altaffilmark{1}, Christopher S. Kochanek\altaffilmark{2}} 

\altaffiltext{1}{Homer L. Dodge Department of Physics and Astronomy,
University of Oklahoma, Norman, OK, 73019;
Rhiannon.D.Griffin-1@ou.edu, xdai@ou.edu}
\altaffiltext{2}{Department of Astronomy and the Center for Cosmology and Astroparticle Physics, 
  Ohio State University, Columbus, OH 43210; ckochanek@astronomy.ohio-state.edu}

\begin{abstract}
Galaxy clusters are predicted to produce $\gamma$-rays through cosmic ray interactions and/or 
dark matter annihilation, potentially detectable by the \fermi\ Large Area Telescope (\fermi-LAT). 
We present a new, independent stacking analysis of \fermi-LAT photon count maps using the 
78 richest nearby clusters ($z<0.12$) from the Two Micron All-Sky Survey (2MASS) 
cluster catalog.  We obtain the lowest limit on the photon flux to date, $2.3 \times 10^{-11}~\pflux$ (95\% confidence) 
per cluster in the 0.8--100~GeV band, which corresponds to a luminosity limit of $3.5\times10^{44}~\plumin$. 
We also constrain the emission limits in a range of narrower energy bands.
Scaling to recent cosmic ray acceleration and $\gamma$-ray emission models, we find that 
cosmic rays represent a negligible contribution to the intra-cluster energy density and gas pressure.  
\end{abstract}


\section{Introduction}

Galaxy clusters are the largest gravitationally bound structures in the universe and as such are important 
tools for studies of structure formation and cosmology.  Past and current methods of detection include 
optical and X-ray observations, the Sunyaev-Zeldovich effect, and gravitational lensing (e.g., Kravtsov \& Borgani 2012 
and references within).  Each new way of observing galaxy clusters can reveal more about these objects, the physics 
involved, and the history of their formation.  

The prevailing structure formation theory suggests that galaxy clusters formed through the hierarchical merging 
of smaller systems, driven by the gravity of the dominant dark matter.  During the merging process, merger
shocks form in the baryons, accelerating cosmic ray (CR) particles to ultra-relativistic speeds with 
Lorentz factors $\Gamma \gg 1000$ \citep[e.g.,][]{volk96, berezinsky97}. Evidence for this can be seen in the form of cluster radio halos or relics, 
spatially extended radio emission or giant radio arcs on scales of $\sim$1~Mpc due to synchrotron emission by CR 
electrons \citep{feretti12}. Electrons lose energy quickly ($\le 10^{8}$~yr) due to the high efficiency of synchrotron emission, 
non-thermal bremsstrahlung, and up-scattering of CMB radiation and so radio relics only probe recent events.  

The shock models also predict that a much larger amount of energy is deposited in the hadronic component 
(ultra-relativistic protons).  CR protons have a very long 
cooling time ($\ge 10^{10}$~yr) and interact with protons in the hot intergalactic medium (1--10 keV) of the 
clusters (e.g., \citealt{berezinsky97,berrington,formofgal}).  Hadronic debris from these p$-$p interactions includes neutral pions, 
whose main decay channel is two \gam-rays: $\pi^0 $ \textrightarrow\ 2\gam\ (99\%) \citep{pions}. 
This $\pi^0$ decay is expected to dominate the CR induced \gam-ray emission which is predicted to
be detectable by the \fermi\ Large Area Telescope and other \gam-ray missions (\fermi-LAT, e.g., Ackermann et al.~2013; 
Vazza \& Br$\ddot{\text{u}}$ggen~2014, Reimer et al.~2003, Aleksi$\acute{\text{c}}$ et al.~2012, Arlen et al.~2012).

Other processes that contribute to the \gam-ray emission are 
inverse Compton scattering and relativistic bremsstrahlung emissions, but these are likely subdominant (e.g., Jeltema et al.\ 2009;  Pinzke \& Pfrommer~2010; Vazza \& Br$\ddot{\text{u}}$ggen~2014; Brunetti \& Jones 2014).
In addition, to detect any dark matter annihilation signal in clusters and set stringent constraints on dark matter annihilation cross-sections, 
the CR emission is a background that must be characterized as part of the spectrum (Ackermann et al.~2013; Huber et al.~2013). 

Several recent papers focus on \fermi-LAT searches for this \gam-ray emission; however, no diffuse \gam-ray emission from galaxy clusters has firmly been detected (Ackermann et al. 2014, Huber et al. 2013, Prokhorov and Churazov 2014 by stacking $\sim50$ clusters; Ackermann et al. 2014, Han et al. 2012, Zandanel and Ando 2014 for individual clusters).
\fermi-LAT has detected point-like \gam-ray emission from the radio galaxies at the centers of the Virgo and Perseus clusters, although these are not attributed to neutral pion decay in the intergalactic medium (Abdo et al. 2009a, 2009b).
In these studies, extragalactic sources beyond the 2FGL catalog\footnote{\texttt{http://heasarc.gsfc.nasa.gov/W3Browse/all/fermilpsc.html}} 
\citep{2fgl} could cause contamination (Han et al. 2012, Ackermann et al. 2014, Prokhorov and Churazov 2014).
The results of the stacking analyses of Ackermann et al. (2014), Huber et al. (2013), and Prokhorov and Churazov (2014) establish the lowest flux upper limits to date and \citet{huber} reached the lowest limits at 2.8--4.9$\times10^{-11}~\pflux$ in the 1--300~GeV band.
Flux upper limits from these stacking analyses are in partial conflict with current models of CR acceleration (e.g., Huber et al.\ 2013).  
Vazza \& Br$\ddot{\text{u}}$ggen (2014) argued that the expected \gam-ray emission for most clusters with 
radio relics should be close to or above the flux limits set by these stacking analyses.  

In this Letter, we present an independent study on this topic, using a uniformly selected sample of nearby clusters.
Our sample is unique among \fermi\ cluster stacking analyses as all of the studies mentioned above use only high X-ray flux HIFLUGCS clusters (Reiprich \& B$\ddot{\text{o}}$hringer~2002). 
Our final selection of 78 clusters includes just 5 HIFLUGCS clusters and of the original 162, 33 are in the HIFLUGCS catalog.
In \S2 we discuss the cluster sample, the $\gamma$-ray data reduction process and the stacking
analysis, and we discuss the results and consequences in \S3.  We assume
$H_0 = 70$~km~s$^{-1}$Mpc$^{-1}$, $k = 0$, $\Omega_M = 0.3$, and $\Omega_\Lambda = 0.7$.

\begin{figure*}[ht]
  \center
  \includegraphics[width=1.0\textwidth]{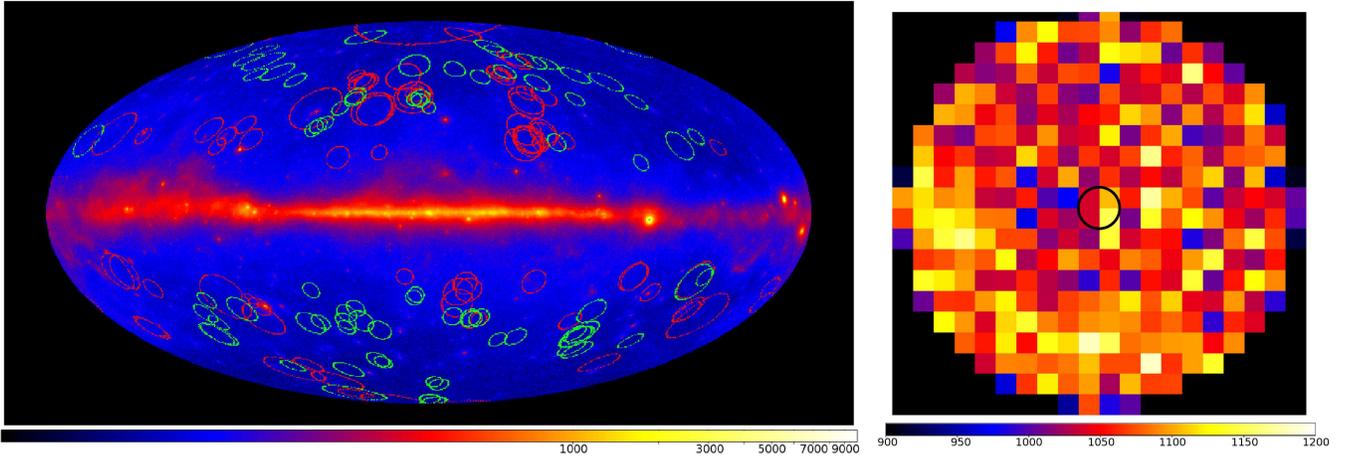}
  \caption{\label{fig:allsky} (Left) All-sky photon count map for the analyzed data. The green ellipses show the locations and area of the 78 rich 2MASS clusters included in the final analysis and red ellipses show the rejected clusters.
   (Right) Final $0.8$--$100$~GeV stacked image for the 78 clusters.  The analysis region is 20~Mpc
    in radius and we limit the flux in the central 2~Mpc source region (black circle). 
    }
\end{figure*}

\section{Analysis} 
\label{sec:thedata}

Our cluster sample consists of the richest, nearby ($z <0.12$) clusters in the 2MASS catalog of clusters identified
using a matched filter algorithm (\citealt{kochanek}).  Like \fermi, 2MASS is also an all-sky survey, 
and we start with the 162 richest clusters outside of the Galactic plane ($|b|>20^\circ$) with $z <0.12$ ($\bar{z}=0.08$) and
 typical masses of $M_{200} \sim 6 \times 10^{14} M_{\odot}$.  
 The sample is well-characterized in richness and distance with extensive calibrations using both near-IR and X-ray
stacking analyses \citep{dai07,dai10, blackburne}.  We eventually use 78 of the original 162 clusters
as discussed in \S3. Details of these clusters are listed in Table~\ref{tab:clusters}.

We downloaded the Pass 7 LAT data from the Fermi Science Support Center (FSSC)\footnote{\texttt{http://fermi.gsfc.nasa.gov/ssc}}, 
along with the Fermi Science Tools (version \texttt{v9r31p1}).  We used the pre-generated weekly all-sky files which span
2008--08--04 to 2013--06--20 for a total of 255 weeks ($\sim5$ years) for SOURCE class photon events.  We followed the 
FSSC Data Preparation
recommendations for our analysis.   Since the point spread function (PSF) of \fermi-LAT decreases with energy, we used a minimum energy 
threshold of $\sim 1$~GeV so that the PSF is always more compact than $0.6$~deg, which also lowers the contributions
of point sources.  A zenith angle cut of 100\degree\ was applied to avoid CR-produced \gam-rays originating from the Earth's atmospheric limb.  
Good time intervals were identified using the recommended selection expression (\texttt{(DATA\_QUAL==1) \&\& (LAT\_CONGIF==1) \&\& ABS(ROCK\_ANGLE)$<$52})
to exclude periods of dead time during spacecraft maneuvers, software updates, and transits through the Southern Atlantic Anomaly.

We first extracted count and exposure maps for each week and then stacked them in time to make a single count and 
exposure map for each cluster.  We then stacked the clusters to obtain the final stacked image.
As we generate the map for each cluster, we search for high background flares from variable $\gamma$-ray sources in the weekly images, 2$\sigma$ above the mean photon flux (\pflux), and
reject these time periods. 
We used seven logarithmically spaced energy bins to cover the  $0.8-100$ GeV band and 
the exposure maps were calculated at the mean energy of each bin. 

Since clusters at higher redshifts have smaller angular sizes, we combine the clusters over 
a fixed 20~Mpc radius region of interest (ROI) binned into 2~Mpc pixels.  This is more
physical than stacking on a fixed angular scale as done previously (Ackermann et al. 2013;
Huber et al. 2013;  Prokhorov \& Churazov 2013).  The cluster emission should lie only
in the central 2~Mpc and the remainder provides the background region.  We also weight
the clusters by $z^2$ so that the stacked signal is not dominated by nearby clusters. This also helps to reduce the variance in the final, stacked image. 
Since the 2~Mpc extraction region can be smaller than the \fermi\ PSFs at lowest energies $\sim 1$~GeV, we calculated energy-dependent aperture flux corrections, and applied them to the flux limits calculated in all energy bands. 

We masked bright sources from the 2FGL catalog \citep{2fgl} using 0\deg5 radius circles to minimize contributions from known point sources to the background. 
This mask radius is larger than the PSF of point sources for all but the lowest energies ($\sim 1$~GeV) considered in our analysis.
We tested various mask sizes and found that 0\deg5 radius resulted in the smoothest background, 
although there is still some contamination to the background, up to 35\% of the source signal.
This contamination contributes randomly to the background and has little effect on the final flux estimates.
The masked region was then statistically filled using the average local background, which we defined to be the annulus with inner and outer radii of 0\deg7 and 0\deg9, respectively.
This average local background contains little contamination from the bright source itself  ($<8\%$ of the source signal).
For multiple bright sources, we masked from brightest to dimmest to minimize the contamination on adjacent masks.
We then visually rejected clusters with poorly masked, bright 2FGL sources.
We flattened each count map to the average exposure near the center of each cluster 
to make the effective exposure time uniform across the image.  For any overlapping
clusters (separation $<2$~Mpc) we excluded the more distant cluster. We also excluded one cluster where a bright source mask completely covers the cluster region. 
These procedures left us with $78$ clusters (Figure~\ref{fig:allsky}, left) with the final stacked 0.8--100~GeV image shown in Figure~\ref{fig:allsky} (right).  
In addition to this map we also examined maps in which we sequentially stack
the clusters in order of increasing background variance.  This ``stacking by
variance'' method provides an alternate approach for images with complex,
multi-component backgrounds including bright sources, the Galactic background, and the diffuse extragalactic
background.

\section{Results and Discussion}

\begin{figure}[t]
   \center
   \includegraphics[width=0.5\textwidth]{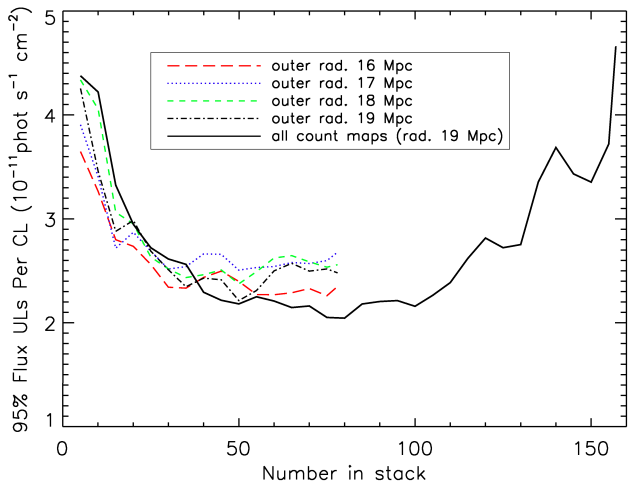}
   \caption{\label{fig:fluxul4} Photon flux upper limits [$0.8-100$ GeV] per cluster as a function of the number of clusters
     included in the stack for different
     outer background radii (the inner radius is fixed at 3~Mpc).   The stacks are ordered by the increasing variance of the cluster maps. The limits initially decline 
     and then flatten.  The black solid line indicates the stacking analysis that includes all count maps not in the Galactic plane, even those with poorly masked 2FGL sources.  The trend increases after $N \sim 100$
     suggesting that we correctly rejected images with large contamination.
}
\end{figure}

Within our 2~Mpc source region of the final stack of 78 count maps (see Figure~\ref{fig:allsky}, right) we detect no excess \gam-ray emission above
the background.  We find an aperture corrected 95\% confidence upper limit 
of $2.48\times 10^{-11}$ \pflux\ per cluster in the $0.8-100$ GeV band when we compare the source
region to 500 
random 2~Mpc regions in our standard background annulus ($3$--$19$~Mpc).  
This Monte-Carlo approach is more general in that it does not assume a Poisson background.
The choice
of the outer background radius has little effect (see Figure~\ref{fig:fluxul4}). 
Figure~\ref{fig:fluxul4} also shows how the limits depend on the number of clusters as we
stack them in order of increasing variance.  There is an initial, rapid decline and
then a flattening with a minimum at $N=45$--$55$ clusters.  
Taking the median of the best upper limits from the stacking by variance method from the four different background apertures,  we obtain the final upper limit of 
 $2.32\times10^{-11}~\pflux$ per cluster corresponding to a luminosity limit of $3.5\times10^{44}~\plumin$ in the $0.8-100$~GeV band
given the median redshift of $z = 0.0758$. If we extend this to include clusters with
poorly masked 2FGL sources (black, solid line in Figure~\ref{fig:fluxul4}), the number of clusters increases to 155 but the limits
begin to significantly worsen as we reach $N>100$ clusters.  This indicates that 
our exclusion of these clusters was well-justified.   
We also constrain the $\gamma$-ray emission upper limits in a range of narrower energy bands and these limits are listed in Table~\ref{tab:fluxul}.

\begin{figure*}[ht]
  \center
  \includegraphics[width=1.0\textwidth]{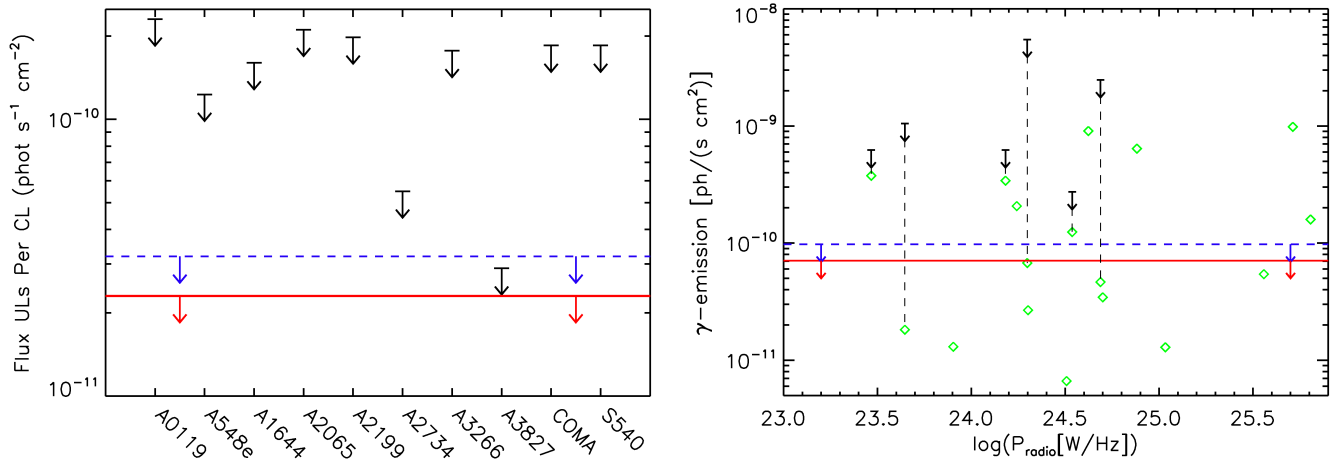}
  \caption{\label{fig:comp} (Left) Flux limits from recent \fermi-LAT studies.  The arrows 
       are 10 clusters randomly selected from \citet{fermi}.  The lower red, solid
       line is our new upper limit, and the upper blue, dashed limit is from \citet{huber}.
   (Right) Model predictions for
       the $0.2$--$100$~GeV flux for galaxy clusters with radio relics (green diamonds, from Vazza~\&~Br$\ddot{\text{u}}$ggen 2014)
       as compared to limits from individual cluster (black arrows, Ackermann~\etal 2013) and the stacking limits from
       Huber et al. (2013, blue, dashed lines) and our analysis (red, solid line). We converted our 0.8--100~GeV limit to the 0.2--100~GeV band by scaling Vazza~\&~Br$\ddot{\text{u}}$ggen's (2014) conversion of Huber et al. (2013) limits that assumed a proton energy index of 2. Notice that the flux limit derived from our stacking method is less than that found by Huber et al. (2013) and therefore the comparison of our upper limit can emphasize the problem described by Vazza~\&~Br$\ddot{\text{u}}$ggen (2014), where upper limits on \gam-ray flux obtained from stacking methods are lower than the predicted fluxes from nearby clusters.
    }
\vspace{0.2in}
\end{figure*}

Figure~\ref{fig:comp} compares our aperture corrected 95\% confidence limits for the $0.8-100$ GeV band 
to the results of \citet{fermi} (1--200~GeV) and \citet{huber} (1--300~GeV). 
We corrected for energy band differences by modeling the photon flux as $dN/dE \propto \left(E/E_0\right)^{-2}$  \citep{pfrommer,huber}, 
but the corrections are very small ($\sim 5\%$).  As seen in Figure~\ref{fig:comp}, our new flux limits are an order of magnitude 
stronger than
those for typical individual clusters and a factor of 2.1--1.2 improvement on the \citet{huber} stacking limits of 2.8--4.9$\times10^{-11}~\pflux$.
Because the \citet{huber} sample is slightly closer, with a mean redshift of $z=0.052$, this results in a factor two difference in $z^2$ compared to our sample. 
The mean mass of
the \citet{huber} clusters is also roughly a factor of two larger at $M_{500} = 5.6\times10^{14}~\msun$, thus our mass-weighted luminosity limit is twice that of \citet{huber}, but for slightly smaller systems.
In the 10--300~GeV band, our mass-weighted luminosity limit is also consistent with the constraint from \citet{prokhorov}.

\begin{deluxetable}{cccccccccccc}

\tabletypesize{\scriptsize}
\tablecolumns{12}
\tablewidth{0pt}
\tablecaption{Galaxy Cluster Sample \label{tab:clusters}}

\tablehead{
\colhead{Name} & \colhead{RA} & \colhead{Decl} & \colhead{z} & \colhead{Richness}
}
\startdata

2MASSCL\ J0330$-$5235   &          52.590 &         $-$52.585 &           0.060 &          28.758 \\
2MASSCL\ J0317$-$4417   &          49.438 &         $-$44.297 &           0.074 &          13.183 \\
2MASSCL\ J0108$-$1526   &          17.228 &         $-$15.435 &           0.053 &          10.392 \\
2MASSCL\ J0327$-$5323   &          51.841 &         $-$53.392 &           0.061 &          16.375 \\
2MASSCL\ J0343$-$5338   &          55.753 &         $-$53.643 &           0.059 &          20.109 \\
2MASSCL\ J0312$-$4725   &          48.191 &         $-$47.417 &           0.081 &          10.006 \\
2MASSCL\ J2235+0129   &         338.935 &           1.496 &           0.059 &          14.214 \\
2MASSCL\ J0112+1611   &          18.103 &          16.196 &           0.061 &          11.029 \\
2MASSCL\ J0544$-$2558   &          86.200 &         $-$25.968 &           0.042 &          15.188 \\
2MASSCL\ J1311+3915   &         197.832 &          39.262 &           0.072 &          15.287 

\enddata
\tablecomments{Position, redshift and richness for the 78 galaxy clusters used in the final stacked image, where the richness is the number of galaxies with luminosity $L > L_*$ \citep{kochanek}.  The clusters are ordered by the variance of the background, starting with the lowest variance cluster.
Note that 5 of the 78 clusters in our final stack are HIFLUGCS clusters (Reiprich \& B$\ddot{\text{o}}$hringer~2002).
  Table~1 is published in its entirety in the electronic edition of the Astrophysical Journal.  A portion is shown here for guidance regarding its form and content.
}
\end{deluxetable}

\begin{deluxetable}{cccccccccccc}

\tabletypesize{\scriptsize}
\tablecolumns{12}
\tablewidth{0pt}
\tablecaption{Photon Flux Upper Limits From Stacking Analyses of Galaxy Clusters\label{tab:fluxul}}

\tablehead{
\colhead{Energy} & \colhead{Outer Bkg.} & \colhead{Flux UL} & \colhead{Lowest} 
\\
\colhead{Range}  & \colhead{Radius}   & \colhead{($N \equiv 78$)}  & \colhead{UL ($N$)}
}
\startdata

$  0.8-100.0$    & 16 Mpc &      23.5   &   22.6      ($N=75 $)  \\
$ 13.0-300.0$    & 16 Mpc &      1.32   &   1.24      ($N=70 $)  \\
$  0.8-  1.6$    & 16 Mpc &      20.9   &   17.2      ($N=70 $)  \\
$  1.6-  3.2$    & 16 Mpc &      8.06   &   7.87      ($N=55 $)  \\
$  3.2-  6.3$    & 16 Mpc &      3.35   &   3.13      ($N=50 $)  \\
$  6.3- 13.0$    & 16 Mpc &      1.88   &   1.83      ($N=70 $)  \\
$ 13.0- 25.0$    & 16 Mpc &      1.01   &   1.01      ($N=78 $)  \\
$ 25.0- 50.0$    & 16 Mpc &      0.591  &    0.569      ($N=70 $)  \\
$ 50.0-100.0$    & 16 Mpc &      0.419  &    0.415      ($N=75 $)  \\
$100.0-170.0$    & 16 Mpc &      0.322  &    0.293      ($N=65 $)  \\
$170.0-300.0$    & 16 Mpc &      0.278  &    0.236      ($N=75 $)	 
\enddata
\tablecomments{Flux upper limits (UL) per cluster at 95\% confidence in units of $10^{-12} $\pflux for various energy ranges. Shown here are our results with an outer background radius of 16 Mpc. 
  First Column: The energy range considered. 
  Second: Outer radius of the background annulus.  
  Third:  Upper limits for the complete stack of 78 clusters.  
  Fourth: Lowest flux upper limits found for the number of clusters producing the best limit in the stacking by variance method.  
}
\end{deluxetable}

Gamma-ray emission from galaxy clusters probes the non-thermal component of the intra-cluster gas.
We can compare our flux limits to recent model predictions for the \gam-ray emission from clusters 
by \citet{huber}, assuming that $\pi^0$ decay is the main \gam-ray emission source.
Since our mass weighted luminosity limit is twice that of \citet{huber},  
we place upper limits of the CR-to-thermal energy ratio of 8\% as scaled to \citet{huber},
corresponding to a CR-to-thermal pressure ratio of $P_{CR}/P_{Th} \simeq 4$\%.  
These results 
confirm the recent claims \citep{huber, prokhorov} that the CR energy and pressure contribute only marginally to the 
total energy and pressure of the intra-cluster gas.
This reduces one uncertainty in estimating the hydrostatic cluster mass using thermal X-ray emission.

Moreover, \gam-ray emission from galaxy clusters provides an additional window to constrain the details 
of cluster formation.  Large-scale cosmological simulations have successfully predicted the mass function 
of galaxy clusters.  However, we lack additional constraints to test the details of these models because 
we generally observe only the final stages of the merging history. Since the cooling time of the hadronic CR component is 
longer than a Hubble time, the hadronic CR component essentially accumulates in clusters 
\citep[e.g.,][]{berezinsky97} so that the final \gam-ray emission produced by the hadronic CRs 
depends on the full merger history.  This should be compared to the CR electron-driven synchrotron
radio emission -- both are driven by the same shocks but the radio emission depends
only on recent activity due to the short CR election life times. 
Since cluster merger models predict that all clusters have experienced similar shocks during the assemblage history, the clusters with radio relics are considered an evolutionary stage of clusters because of the fast electron cooling time-scale.

Thus, we can compare the \gam-ray flux constrained from a general cluster population with the radio flux from clusters with radio relics, because both the CR components are accelerated by the same shocks.
While some model predicted \gam-ray fluxes from clusters with radio halos are consistent with our limits (e.g., Kushnir et al.\ 2009),
 using a semi-analytical model, 
Vazza \& Br$\ddot{\text{u}}$ggen (2014) calculated the expected \gam-ray emission from clusters with radio relics (arcs)
that are evidence for shocks with Mach numbers of $2$--$4$. 
Figure~\ref{fig:comp} (right) compares
the predictions of Vazza \& Br$\ddot{\text{u}}$ggen (2014) (green diamonds) to the observed limits (arrows and horizontal lines), and, like Huber et al. (2013) our limits are
well below the typical predictions. 
The problem can be more severe because there can be multiple mergers as a cluster forms.
Several papers explored scenarios to resolve this discrepancy,
including over-estimated Mach numbers from the radio data, lower energy deposition rates to the hadronic 
CR component than in standard diffuse shock acceleration models, and reacceleration of electrons (e.g., Vazza \& Br$\ddot{\text{u}}$ggen 2014; Brunetti \& Jones 2014; Zandanel et al.\ 2014).  

\acknowledgements
This research has made use of the publicly available \fermi-LAT data.
We thank the anonymous referee and G.\ Brunetti, T.\ A.\ Thompson, D.\ Kushnir, B.\ Katz, F.\  Zandanel, and J.\ Han for helpful comments.

\end{document}